# Gate-controlled ultraviolet photo-etching of graphene edges

Nobuhiko Mitoma, and Ryo Nouchi[a]

Nanoscience and Nanotechnology Research Center, Osaka Prefecture University, Sakai, Osaka 599-8570, Japan

The chemical reactivity of graphene under ultraviolet (UV) light irradiation is investigated under positive and negative gate electric fields. Graphene edges are selectively etched when negative gate voltages are applied, while the reactivity is significantly suppressed for positive gate voltages. Oxygen adsorption onto graphene is significantly affected by the Fermi level of the final state achieved during previous electrical measurements. UV irradiation after negative-to-positive gate sweeps causes predominant oxygen desorption, while UV irradiation after gate sweeps in the opposite direction causes etching of graphene edges.

___________________________

[a] Author to whom correspondence should be addressed. Electronic mail: r-nouchi@21c.osakafu-u.ac.jp, Telephone/Fax: +81 72 254 8394

Single layer graphite, graphene, shows extraordinarily high charge carrier mobility that has been exploited to develop high-frequency transistors[1] and ultrafast photodetectors.[2] In addition, graphene is an ideal two-dimensional material that has the highest surface-to-volume ratio, which enables it to be extremely sensitive to adsorption of surrounding gas molecules, *i.e.*, to be suitable for sensing application.[3] In particular, oxygen molecules, which are ubiquitous in the air, are known to have a significant effect on the electrical properties of graphene, e.g., through p-type charge carrier doping.[4-8] This phenomenon is understood to be the result of electron transfer from graphene to physisorbed oxygen. Desorption of such oxygen adsorbates was reported to occur under ultraviolet (UV) light irradiation.[3,9] UV irradiation has also been adopted to reduce the electrode-contact resistance;[10] however, despite the usefulness of the UV treatment, the results obtained in former studies have varied from group to group. Several groups reported that short UV exposure undopes graphene,[3,4] while others have reported that UV exposure causes p-[7] or n-type[11] doping. Long-term UV exposure is known to cause heavy p-type doping and/or C-C bond reconstruction from $sp^2$ to $sp^3$.[12,13] However, there remains a lack of systematic understanding of the UV-induced photochemical reactions in graphene. It was reported that the adsorption of oxygen molecules onto bilayer graphene can be controlled through tuning of the Fermi level with a gate electric field.[6] In this study, we focus on the chemical reaction between graphene and oxygen caused by UV irradiation, and the physisorption of oxygen molecules when a gate voltage is applied to graphene field effect transistors (FETs).

Single-layer graphene flakes were mechanically exfoliated onto a 300-nm-thick thermally oxidized silicon layer on a heavily p-doped silicon substrate using an adhesive tape.[14] The number of graphene layers was confirmed using the optical contrast and Raman spectra.[15] Electron beam lithography was performed to pattern electrodes, followed by the formation of Cr (as an adhesion layer) and Au thin films (1 and 60 nm thick, respectively) using vacuum (ca. $10^{-4}$ Pa) deposition, and a lift-off process. The fabricated graphene FETs were exposed to UV light (Ushio Inc., UXM-Q256BY, 1 W cm$^{-2}$; the spot size was ca. 0.5 mm in diameter and was sufficiently larger than the graphene flakes) in a probe station system (Fig. 1). All the experiments were conducted in air at room temperature (ca. 295 K). The relative humidity of the experimental environment was approximately 45%.

Raman spectroscopy measurements were performed using a laser Raman microscope system (Nanophoton Corporation, RAMAN-DM). Figure 2(a) shows a false color image of the obtained Raman $D$ (ca. 1340 cm$^{-1}$) and $2D$ (ca. 2680 cm$^{-1}$) intensities, before and after UV irradiation with the gate voltage $V_G$ at -60 V and the source-drain voltage $V_{SD}$ at 1 mV. These two characteristic peaks originate from double resonant Raman scattering processes near the **K** points in the Brillouin zone. The $D$ band is a combination of the **K** point phonon of $A_{1g}$ symmetry and elastic scattering by defects. The $A_{1g}$ mode is forbidden in perfect honeycomb sp$^2$ carbon networks; therefore, the $D$ band is known to be indicative of structural disorder or chemical modification of the graphene planes.[16] The $2D$ band is due to two phonons with opposite momentum in the highest optical branch near the **K** point. The energy dispersion of graphene is significantly different depending on the number of graphene layers, i.e., the unit cell structure of the material. The difference between the sharp Gaussian-like spectrum for single-layer graphene and the four-component spectrum for bilayer graphene is explained by the difference in the energy band structure.[15] The sharp Gaussian-shaped $2D$ band that appeared for the entire graphene sheet confirms that it is single-layer graphene (Fig. 2). At the initial state, the obtained single-layer graphene was highly crystalline, and a negligibly small $D$ band was observed at the center of the graphene and even at the edge (Fig. 2(b)). It should be noted that the laser used in the present study is not polarized; therefore, the $A_{1g}$ modes at the edge should not necessarily be detected due to the random edge structure.[17] However, after UV irradiation for 20 min under application of $V_G$ at -60 V and $V_{SD}$ at 1 mV, a strong $D$ band appeared around the graphene edge (Figs. 2(a), 2(c)). UV irradiation with application of a negative gate voltage was conducted for four graphene FETs and all exhibited evidence of selective etching from the graphene edges. The Raman spectra obtained from the edge region after UV irradiation had broad $D$ and $D'$ (ca. 1600 cm$^{-1}$) bands. Such peaks have been reported in reduced graphene oxide.[18] The loss of the $G$ (ca. 1580 cm$^{-1}$) and $2D$ bands indicates that the hexagonal lattice structure around the graphene edge was destroyed and amorphized.[8,16] Thus, we refer this reaction to etching, not to modification.

Interestingly, such selective etching from the edges was not observed for application of $V_G$ at +60 V and $V_{SD}$ at 1 mV during UV irradiation. None of four tested graphene FETs showed a $D$ band after UV irradiation under positive gate voltage conditions. These results indicate that the photochemical reactivity is strongly dominated by the surface polarity and/or the Fermi level of graphene.[6,19,20] The binding energy for graphene with oxygen or ozone molecules has been reported to be ca. 0.2 or 0.25 eV.[21,22] The tuning range of the Fermi level for graphene by the application of $V_G$ at ±60 V through a 300 nm thick SiO$_2$ layer is ca. 0.24 eV, which should be sufficient to change the oxygen/ozone adsorption capability. The Fermi level shift may heighten or lower the energy barrier for the physisorption of oxygen. The etching reaction is considered to occur subsequent to the conversion from low-reactivity oxygen to high-reactivity ozone, which is triggered by UV irradiation. The conversion efficiency is affected by the binding energy between graphene and physisorbed oxygen molecules.

Selective functionalization of the graphene edge using mild NH$_3$ plasma was previously reported.[23] The high chemical reactivity at the edges is now attributed to the presence of dangling bonds.[24] Graphene modification was observed only at the edge in the case of mild plasma treatment. In contrast to this, a 3-µm-wide graphene flake was completely etched away during the 20 min UV irradiation in the present study. This indicates that the etching process is not restricted to the edges but proceeds into the center regions.

Oxygen molecules adsorbed under an applied gate voltage remain adsorbed on graphene for at least several hours at room temperature.[6] This means that the photochemical reactivity caused by UV light treatment is strongly influenced by the final state that results from previous electrical measurements. The photochemical reactivity of the graphene by UV irradiation was investigated immediately after the measurement of transfer curves (gate voltage $V_G$, dependency of the conductivity, $\sigma$). Transfer curves for graphene FETs can be obtained by sweeping the gate voltage from negative to positive values or vise versa; therefore, the photoreactivity should depend on the gate sweep direction. The following two cases were examined: the gate voltage was swept from -60 to +60 V, and from +60 to -60 V, which are referred to as the forward sweep and the backward sweep, respectively. $V_{SD}$ during these measurements was set to 1 mV. At the final moment of the forward (backward) sweep, a gate voltage of +60 V (-60 V) was applied to the FETs; thus the final state of the graphene Fermi level is dependent on the gate voltage sweep direction.

The gate voltage sweeps were repeated 10 times. After 10 measurement cycles, UV light with an intensity of 1 W cm$^{-2}$ was used to irradiate the graphene surface for 5 min from a distance of 1 cm. It should be noted that the gate and source-drain voltages were not applied during UV irradiation, unlike for the previous experiment results shown in Fig. 2. This experimental procedure was repeated 7 times, i.e., the evolution of the transfer curve was monitored up to a total UV irradiation time of 30 min. The dotted black curve in Fig. 3(a) is the transfer curve obtained for the initial state. A conventional V-shaped transfer curve that reflected the energy band structure of graphene was observed. The minimum conductivity point, i.e., the Dirac point $V_{Dirac}$, appeared around $V_G$ = +14 V. The non-zero $V_{Dirac}$ is attributable to the measurement being conducted in air. The adsorbed oxygen/water redox couple, $O_2 + 4H^+ + 4e^- \rightleftarrows 2H_2O$ is known to cause p-type doping, where oxygen molecules are stabilized by water solvation.[4,5] The solid black curve in Fig. 3(a) is the transfer curve for the tenth forward sweep. The difference between the first and tenth run is that $V_{Dirac}$ is shifted toward the positive gate voltage direction. $V_{Dirac}$ for the tenth run appeared around $V_G$ of +20 V. Figure 3(b) shows the evolution of $V_{Dirac}$ for the forward sweep cycles. $V_{Dirac}$ shifted toward higher $V_G$ with repetition of the sweep. Such a shift of the Dirac point is explained in terms of the physisorption of oxygen/water on graphene. Repeated gate voltage application is considered to attract the oxygen/water redox couple in air. A reduction in charge carrier mobility was observed simultaneously, as shown in Figs. 3(a) and 3(c), because adsorbed oxygen molecules scatter the charge carriers in graphene.[6] The hole mobility $\mu_h$ was fitted using the following equation:

$$\sigma = -\mu_h C_G (V_G - V_{Dirac}) + \sigma_{Res}, \tag{1}$$

where $C_G$ is the gate capacitance per unit area, which is $1.15 \times 10^{-4}$ F m$^{-2}$ for graphene on 300-nm-thick $SiO_2$, and $\sigma_{Res}$ is the residual conductivity.[25] The data were fitted at points not too close to $V_{Dirac}$, to avoid the influence of charge inhomogeneity.[26] The trends observed in Figs. 3(b) and 3(c) also support the promotion of oxygen/water adsorption by repeated forward sweeps.

In the case of the forward sweep, a $V_G$ of +60 V was applied at the final moment of measurement, i.e., the Fermi level was placed above the Dirac point. After 10 forward sweeps, UV irradiation caused $V_{Dirac}$ to be reduced to a $V_G$ of around +2 V (Fig. 3(b)) and $\mu_h$ to be increased to 0.76 m$^2$ V$^{-1}$ s$^{-1}$ (Fig. 3(c)). The recovery of $V_{Dirac}$ and $\mu_h$ indicate the desorption of oxygen molecules from the graphene plane. This can be explained in terms of oxygen adsorption by gate voltage application and desorption by UV irradiation.

Next, backward sweep cycles followed by UV exposure without a bias voltage were investigated. The results are shown in Figs. 3(d-f). The principal difference between the forward and backward sweeps is evident from the shifts of both $V_{Dirac}$ and $\mu_h$. The upward shift of $V_{Dirac}$ in Fig. 3(e) indicates that the oxygen molecules are strongly pinned on the graphene, even after UV irradiation. The large decrease in $\mu_h$ at the first UV irradiation also indicates that the graphene was altered by oxygen molecules. A slight increase in $\mu_h$ was observed after the second UV irradiation; however, the decreasing trend of $\mu_h$ is stronger for the backward sweep (Fig. 3(f)) than for the forward sweep (Fig. 3(c)). The reaction appears to be caused by physisorbed oxygen, which is converted to reactive ozone under UV irradiation. However, hystereses, the shape difference of transfer curves between the forward and backward sweeps, were observed, which indicates the existence of water around graphene.[27] Such water molecules are considered to support the physisorption of oxygen molecules and may react with graphene under UV irradiation. Further study is required to clarify the roles of oxygen and water molecules.

We have only focused on the dependence of $V_G$ until here. $V_{SD}$ is also an important parameter for the UV-induced photochemical reaction. Raman spectra measured at the graphene edge after the backward sweep cycle with the two different $V_{SD}$ values are shown in Fig. 4. For a $V_{SD}$ of 1 mV, no detectable $D$ band was observed at the edge; however, an obvious $D$ band was observed when a $V_{SD}$ of 10 mV was applied. This may be due to an increase in temperature by Joule heating in the graphene devices, which should activate the chemical reaction. Chemical etching is suppressed near the Au electrodes, as shown in Fig. 2(b), which is attributable to heat dissipation to the conductive Au electrodes.

Comparison of the Raman spectra obtained from the graphene edge in Fig. 2(c) and Fig. 4 indicates that the extent of UV etching of the graphene edge is more significant for the sample shown in Fig. 2(c), even though the applied drain voltage was smaller. This difference in reactivity can be understood in terms of oxygen adsorption that occurs under an applied gate voltage. $V_{SD}$ is smaller in Fig. 2(c); however, a gate voltage was applied during UV irradiation, whereas UV irradiation without gate voltage application provided the result in Fig.

4. Thus, the total amount of consumed oxygen molecules during the etching reaction should be larger in Fig. 2(c).

In summary, selective etching of graphene edges under UV exposure was observed by application of a negative gate electric field. The UV photoreactivity of graphene is significantly influenced by the Fermi level. The reactivity is easily changed after conventional transfer curve measurements, and thus the present findings should be considered for the photochemical treatment of graphene. This etching method can be easily applied to a variety of graphene FET systems and would be a useful microfabrication technique for graphene devices.

This work was supported in part by Special Coordination Funds for Promoting Science and Technology from the Ministry of Education, Culture, Sports, Science and Technology of Japan; and a technology research grant from JFE 21st Century Foundation. N. Mitoma thanks T. Tanimoto for useful advice on the FET measurements. The graphite flake used in the present study was supplied by T. Takano.

[27] H. Wang, Y. Wu, C. Cong, J. Shang, and T. Yu, ACS Nano **4**, 7221 (2010).

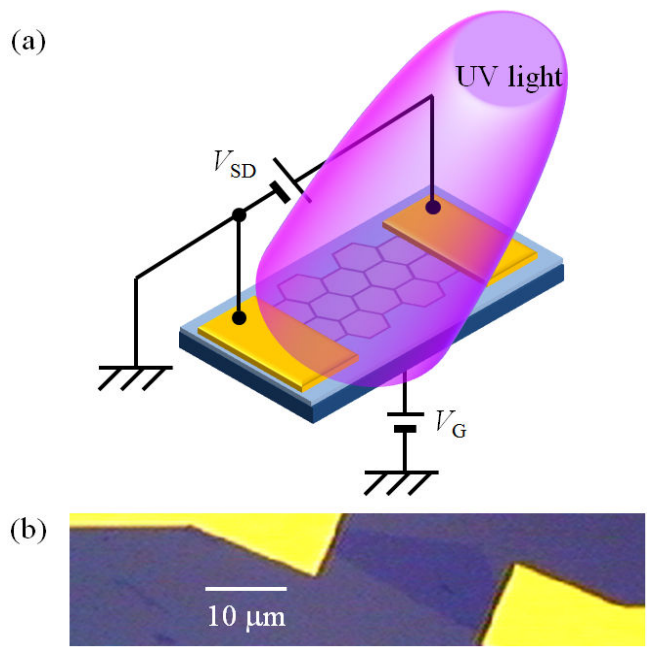

FIG. 1. (a) Schematic image of experimental setup. The graphene sheet was irradiated by UV light with an intensity of 1 W cm$^{-2}$ in air from a distance of about 1 cm. $V_G$ and $V_{SD}$ represent the gate and source-drain voltages, respectively. (b) Optical micrograph of a graphene FET used in the present study.

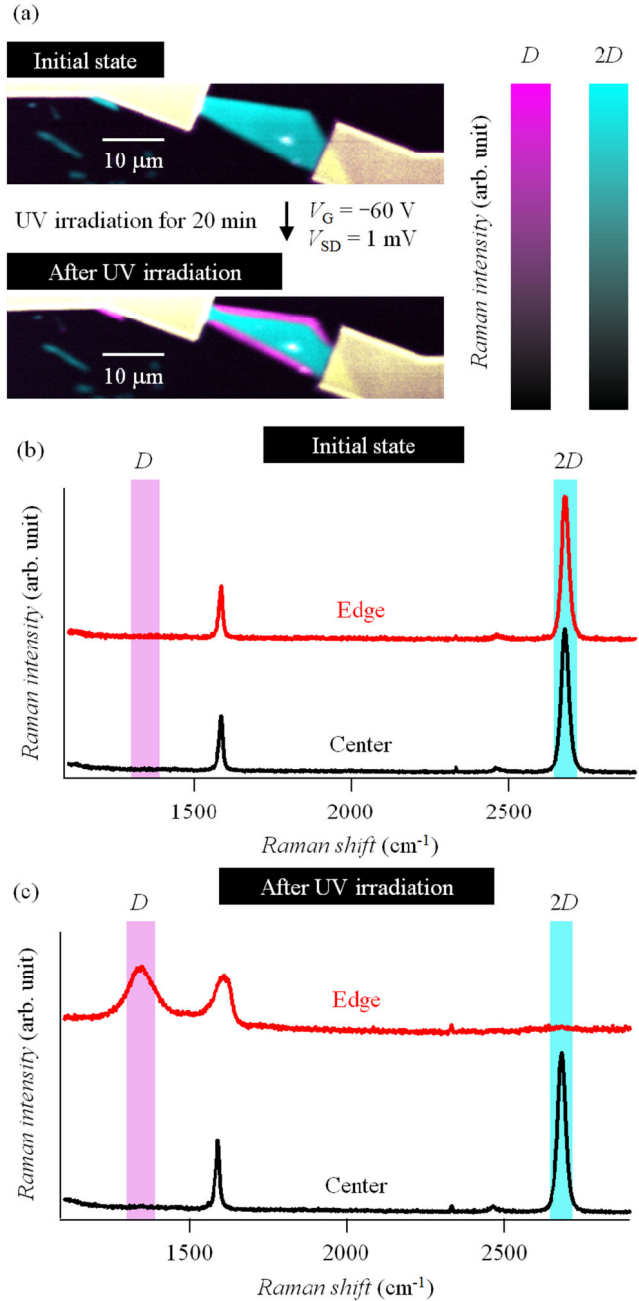

FIG. 2. (a) False color image of the Raman intensity before and after UV irradiation for 20 min with application of a gate voltage of -60 V and a source-drain voltage of 1 mV. Magenta and cyan represent the Raman $D$ and $2D$ band intensities of the graphene, respectively. The yellow areas are the Au electrodes and the dark area is the $SiO_2$ substrate. The white spot apparent in the basal plane of the graphene sheet may be a contaminant created during the device fabrication process. (b,c) Raman spectra obtained from the center and edge of the graphene FET (b) before and (c) after UV irradiation. The spectra are shifted for clarity.

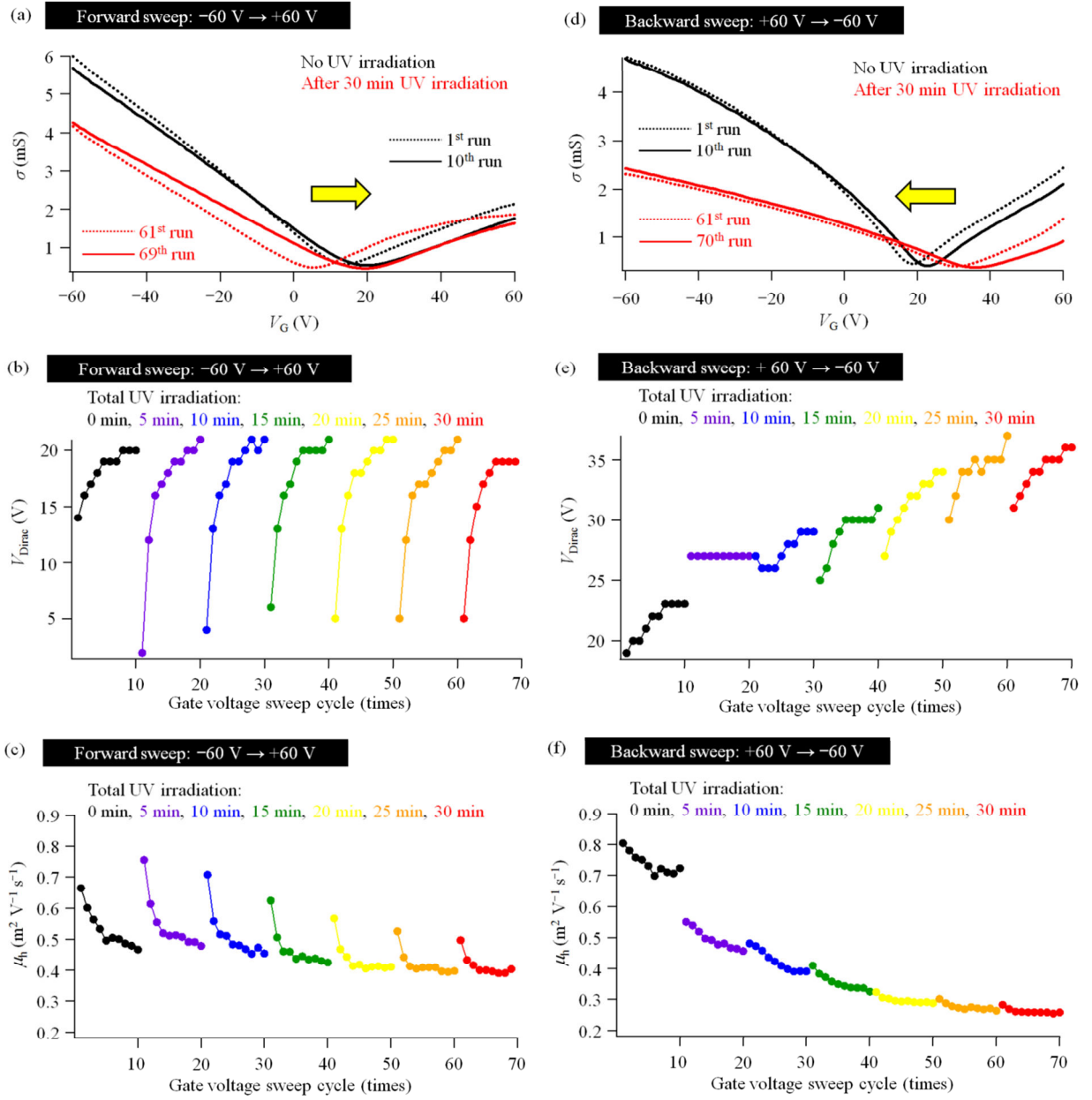

FIG. 3. (a) Evolution of the transfer curve when the gate voltage sweep cycle was repeated at the initial state and after 30 min UV irradiation. Shifts of (b) the Dirac point voltage and (c) the charge-carrier mobility during the gate voltage sweep cycle. To obtain the (a), (b), and (c) data, the gate voltage was swept from -60 V to +60 V. (d), (e), and (f) are the data that correspond to (a), (b), and (c) with the gate voltage swept in the opposite direction. $V_{SD}$ was set to 1 mV during the gate voltage sweeps.

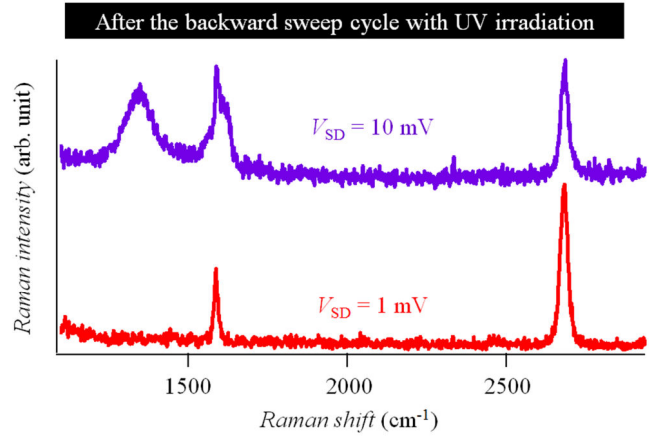

FIG. 4. Raman spectra measured at the graphene edge after the backward sweep cycle with two different $V_{SD}$. UV light was irradiated for a total of 30 min in both cases.